\documentclass[aps,prl,reprint,superscriptaddress,10pt,showpacs]{revtex4-1}
\usepackage{amsmath,amssymb,amsfonts}
\usepackage{graphicx}
\usepackage{txfonts}
\usepackage{color}

\usepackage[unicode,breaklinks]{hyperref}
\hypersetup{
    unicode=true,
    a4paper=true,
    plainpages=false,
    pdftitle={Onsager--Kraichnan Condensation in Decaying Two-Dimensional Quantum Turbulence},
    pdfauthor={T. P. Billam, M. T. Reeves, B. P. Anderson, and A. S. Bradley},
    pdfsubject={Onsager--Kraichnan Condensation in Decaying Two-Dimensional Quantum Turbulence},
    colorlinks=true,
    linkcolor=blue,
    citecolor=blue,
    filecolor=black,
    urlcolor=blue
}
\newcommand{\eqnrefp}[1]{{[Eq.~(\ref{#1})]}}

\newcommand{\eqnreft}[1]{{Eq.~(\ref{#1})}}
\newcommand{\eqnreftfull}[1]{{Equation~(\ref{#1})}}

\newcommand{\figreft}[2]{Fig.~\ref{#1}#2}

\newcommand{\figreftfull}[2]{Figure~\ref{#1}#2}
\newcommand{\figrefp}[2]{[Fig.~\ref{#1}#2]}

\newcommand{\ww}{\mathbf{w}}
\newcommand{\kk}{\mathbf{k}}
\newcommand{\rr}{\mathbf{r}}
\newcommand{\vvv}{\mathbf{v}}

\newcommand{\atan}{\mathrm{atan}}

\begin{document}
\title{Onsager--Kraichnan Condensation in Decaying Two-Dimensional Quantum Turbulence}

\author{T. P. Billam}
\thanks{Author to whom correspondence should be addressed}
\email[Email: ]{thomas.billam@otago.ac.nz}
\affiliation{Jack Dodd Centre for Quantum Technology, Department of Physics,
University of Otago, Dunedin 9016, New Zealand}
\author{M. T. Reeves}
\affiliation{Jack Dodd Centre for Quantum Technology, Department of Physics,
University of Otago, Dunedin 9016, New Zealand}
\author{B. P. Anderson}
\affiliation{College of Optical Sciences, University of Arizona, Tucson, AZ 85721, USA}
\author{A. S. Bradley}
\thanks{Author to whom correspondence should be addressed}
\email[Email: ]{ashton.bradley@otago.ac.nz}
\affiliation{Jack Dodd Centre for Quantum Technology, Department of Physics,
University of Otago, Dunedin 9016, New Zealand}

\date{\today}


\begin{abstract}
Despite the prominence of Onsager's point-vortex model as a statistical
description of 2D classical turbulence, a first-principles development of the
model for a realistic \emph{superfluid} has remained an open problem.  Here we
develop a mapping of a system of quantum vortices described by the homogeneous
2D Gross-Pitaevskii equation (GPE) to the point-vortex model, enabling
Monte-Carlo sampling of the vortex microcanonical ensemble. We use this
approach to survey the full range of vortex states in a 2D superfluid, from the
vortex-dipole gas at positive temperature to negative-temperature states
exhibiting both macroscopic vortex clustering and kinetic energy condensation,
which we term an \textit{Onsager-Kraichnan condensate} (OKC). Damped GPE
simulations reveal that such OKC states can emerge dynamically, via aggregation
of small-scale clusters into giant OKC-clusters, as the end states of decaying
2D quantum turbulence in a compressible, finite-temperature superfluid.  These
statistical equilibrium states should be accessible in atomic Bose-Einstein
condensate experiments.
\end{abstract}

\maketitle

The importance of the point-vortex model as a \textit{statistical} description
of two-dimensional (2D) classical hydrodynamic turbulence was identified by
Onsager \cite{Onsager1949}, who predicted that the bounded phase-space of a
system of vortices implies the existence of negative-temperature states
exhibiting clustering of like-circulation vortices \cite{Eyi2006.RMP78.87}.
This model provides great insight into 2D classical  turbulence
(CT)~\cite{Boffetta12a}, and much subsequent work has focused on the
point-vortex model as an approximate statistical description of decaying 2DCT
\cite{Tabeling2002, Mon1974.PF6.1139, montgomery_etal_pf_1992, miller_prl_1990,
robert_sommeria_jfm_1991}. While classical fluids cannot directly realize the
point-vortex model, atomic Bose-Einstein condensates (BECs) --- which present
an emerging theoretical \cite{Parker05a, Nazarenko07a, Horng09a, Numasato10a,
Numasato10b, Sasaki2010,White10a, Nowak2011a, Nowak2012a, Schole2012a,
Bradley2012a, White2012a, kusumura_etal_jltp_2013,Tsubota2013,
reeves_etal_prl_2013} and experimental \cite{Neely10a, neely_etal_prl_2013, Wilson2013}
paradigm system for the study of quantum vortices and 2D quantum turbulence
(2DQT) --- offer the possibility of physically realizing Onsager's
negative-temperature equilibrium states.  A concrete realization of the
point-vortex model in an atomic superfluid will broaden our understanding of
the universality of 2D turbulence by enabling new studies of spectral
condensation of energy at large scales~\cite{kraichnan_jfm_1975,
Kra1980.RPP5.547, Chertkov2007a, xia_etal_prl_2008}, statistical mechanics of
negative-temperature states \cite{weiss_mcwilliams_pf_1991,
campbell_oneil_jsp_1991, sano_etal_jpsj_2007}, the dynamics of macroscopic
vortex clustering \cite{yatsuyanagi_etal_prl_2005}, and the inverse energy
cascade \cite{SIGGIA1981a,Kra1967.PF10.1417, Lei1968.PF11.671,
Bat1969.PF12.II233}, previously confined to 2DCT.

In this letter we develop an analytic statistical description of the
microstates of 2D quantum vortices within the homogeneous Gross-Pitaevskii
theory, and show that macroscopically clustered vortex states emerge from
small-scale initial clustering as end products of decaying 2DQT.  As in CT, the
homogeneous system offers the clearest insight into the underlying physics, and
is increasingly relevant experimentally~\cite{Gaunt:2013ip}. Consequently, our
results describe physics relevant to a wide range of possible vortex
experiments in atomic BECs. By systematically sampling the microcanonical
ensemble for the vortex degrees of freedom, we give a detailed, unifying view
of the properties of vortex matter in a homogeneous 2D superfluid. We
characterize the emergence of macroscopic clusters of quantum vortices at
negative temperatures, linked with spectral condensation of energy at the
system scale. In the context of 2DQT we call this the
\textit{Onsager--Kraichnan condensate} (OKC), as it represents a physically
realizable state that unifies Onsager's negative-temperature point-vortex
clusters with the spectral condensation of kinetic energy predicted in 2DCT by
Kraichnan \cite{kraichnan_jfm_1975}. 

Atomic BECs support quantum vortices subject to thermal and acoustic
dissipative processes that may be detrimental to the observation of an OKC. We
assess the accessibility of the excited states comprising an OKC via dynamical
simulations according to the damped Gross-Pitaevskii equation (dGPE). We find
that our statistical approach describes the end-states of decaying 2DQT that
emerge dynamically from low-entropy initial states. Even for relatively small
positive point-vortex energies, the OKC emerges as a result of statistically
driven transfer of energy to large length scales.  

\begin{figure*}
\centering
\includegraphics[width=0.95\textwidth]{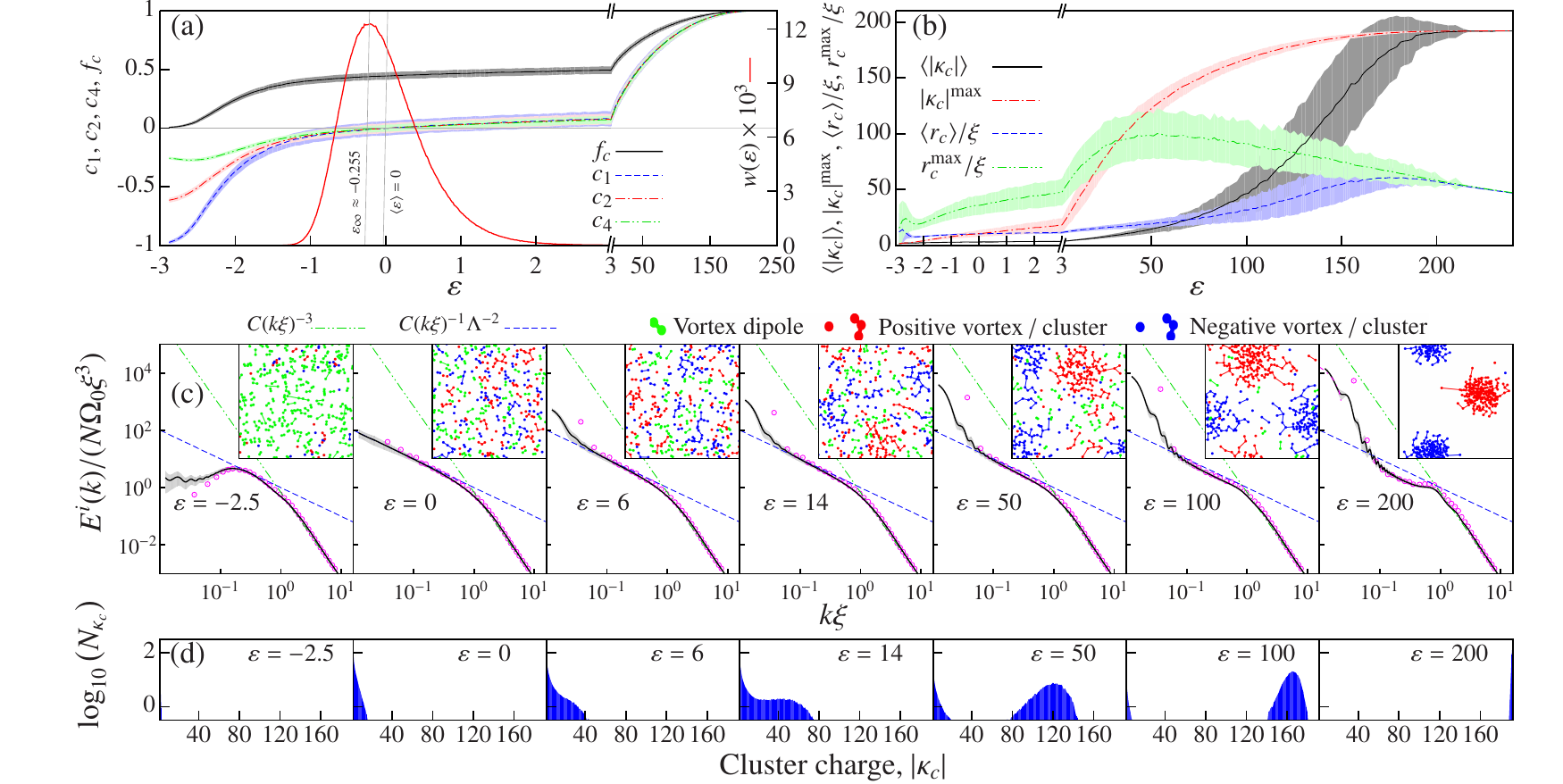}
\caption{(color online). Properties of neutral $N$-vortex states in statistical
equilibrium, corresponding to the end-states of decaying 2DQT.  (a): Clustering
measures $f_c$ and $c_B$ and structure function $w(\varepsilon)$; vertical grey
lines indicate the expected energy ($\langle \varepsilon \rangle$) and the
boundary between positive- and negative-temperature states
($\varepsilon_\infty$).  (b): Average and maximum cluster charges
($|\kappa_c|$) and radii ($r_c$) obtained using the RCA (see text).  (c): IKE
spectrum in the point-vortex-like approximation \eqnrefp{eqn:2dqt_spectrum}
(solid black line, grey shaded area), and obtained from the GPE wavefunctions
\cite{Bradley2012a,Numasato10b} (magenta circles).  Straight lines show
analytical $k^{-3}$ and $k^{-1}$ power laws (see text).  Inset: typical vortex
configurations (field of view $L \times L$): dots indicate vortices, which have
been sorted into dipoles, free vortices and clusters by the RCA (see legend).
Lines show the minimal spanning tree of clusters and identify dipoles.  (d): Distribution of
$N_{\kappa_c}$.  Shaded areas behind curves in (a,b,c) indicate the width of
the equilibrium distribution ($\pm 1$ standard deviations).  Ensemble sizes are
given in \cite{sampling}.
\label{figure_Sampling}}
\end{figure*}

To map the Gross-Pitaevskii theory to the point-vortex model, we introduce an ansatz wavefunction for $N$ vortices in a homogeneous periodic square BEC of side $L$, with
positions $\rr_j$ and circulations $h\kappa_j/m$ defined by charges $\kappa_j=\pm 1$ ($\sum_{j=1}^N \kappa_j = 0$),
\begin{equation}
\psi(\rr,\{\rr_j\},\{\kappa_j\}) = e^{i \theta(\rr,\{\rr_j\},\{\kappa_j\})} \prod_{p=1}^N \chi \left( |\rr-\rr_p| \right)\,,
\label{eqn:ansatz}
\end{equation}
where $\chi(r)$ is the radial profile of an isolated quantum vortex core,
obtained numerically \cite{Bradley2012a}.  Unlike the velocity field for
point-vortices in a doubly-periodic domain
\cite{campbell_oneil_jsp_1991,weiss_mcwilliams_pf_1991}, the associated quantum
phase $\theta$ does not, to our knowledge, appear in the literature. We present
an expression for $\theta$ as a rapidly convergent sum, obtained from a poorly
convergent sum over periodic replica vortices, in the Supplemental Material
\cite{suppinf}. The phase $\theta$ yields a periodic superfluid velocity $\mathbf{v}=(\hbar/m)\nabla
\theta$ very close to the point-vortex velocity, but consistently modified by the boundary conditions such that $\theta (x+\eta_x L,y+\eta_y L) = \theta(x,y) + \zeta
2\pi$, for all $\eta_x,\eta_y,\zeta \in \mathbb{Z}$, remains a well-defined quantum phase. To ensure that $\chi(r)$ is accurate, we enforce a
minimum-separation constraint $|\rr_p-\rr_q|\ge 2 \pi \xi$, where $\xi = \hbar
/ \sqrt{\mu m}$ is the healing length (for BEC chemical potential $\mu$ and
atomic mass $m$), and $1 \le p \ne q \le N$.

Up to an additive constant, the total kinetic energy in the point-vortex model
is $N \Omega_0 \xi^2 \varepsilon(\{\rr_j\},\{\kappa_j\})$. Here, the
dimensionless point-vortex energy (per vortex) is given by
\cite{campbell_oneil_jsp_1991}
\begin{equation}
\varepsilon(\{\rr_j\},\{\kappa_j\}) = \frac{1}{N} \sum_{p=1}^{N-1} \sum_{q=p+1}^N \kappa_p \kappa_q f\left( \frac{\rr_p-\rr_q}{L} \right)\,,
\label{eqn:pv_energy}
\end{equation}
where $f(\rr) \equiv f(x,y) = 2\pi[|y|(|y|-1)+1/6] - \log \{ \prod_{s=-\infty}^\infty
[1 - 2\cos(2\pi x)\exp(-2\pi|y+s|) + \exp(-4\pi |y+s|)] \}$, and $\Omega_0=\pi
\hbar^2 n_0 /m\xi^2$ is the unit of enstrophy for 2D homogeneous superfluid
density $n_0$~\cite{Bradley2012a}.  Accounting for compressible effects with the core ansatz
$\chi_\Lambda(r) = [n_0 r^2/(r^2+\xi^2\Lambda^{-2})]^{1/2}$, the incompressible
kinetic energy (IKE) spectrum of \eqnreft{eqn:ansatz} at scales below the
system size $L$ is well approximated by \cite{suppinf, Bradley2012a}
\begin{equation}
  E^i(k) = \Omega_0 \xi^3 F_\Lambda (k\xi) \left[N + 2\sum_{p=1}^{N-1} \sum_{q=p+1}^N \kappa_p \kappa_q J_0(k|\rr_p-\rr_q|) \right]\,,
\label{eqn:2dqt_spectrum}
\end{equation}
where $F_\Lambda(k\xi) = \Lambda^{-1} g(k\xi \Lambda^{-1})$, $g(z) = (z/4) [
I_1(z/2) K_0(z/2) - I_0(z/2) K_1(z/2)]$, $\Lambda =\xi^2 n_0^{-1/2} d\chi(0)/dr
\approx 0.825$ and $J_\alpha$ ($I_\alpha$, $K_\alpha$) are (modified) Bessel
functions.  \eqnreftfull{eqn:2dqt_spectrum} leads to a universal ultraviolet
(UV, $k\gg\xi^{-1}$) $k^{-3}$ power-law, $E^i(k) = C(k\xi)^{-3}$, where $C=
\Lambda^2 {N\Omega_0 \xi^3}$ \cite{Bradley2012a, Nowak2012a}.  In the infrared
(IR, $k \lesssim \xi^{-1}$) the average spectrum of $N$-vortex configurations
with randomly distributed vortices is equal to the sum of $N$ independent
single-vortex spectra, giving the $k^{-1}$ power-law $E^i(k) = C (k\xi)^{-1} /
\Lambda^2$ \cite{kusumura_etal_jltp_2013, Bradley2012a, Nowak2012a}.

The wavefunction $\psi$ \eqnrefp{eqn:ansatz} is set entirely by
the vortex configuration, allowing us to adopt a statistical treatment where,
for each configuration ($\{\rr_j\}$,$\{\kappa_j\}$), $\psi$ defines a microstate of the 2D BEC~\cite{2dbec}.  Aside
from the minimum-separation constraint, the phase $\theta$ in Eq.~(\ref{eqn:ansatz}) establishes a one-to-one
correspondence between the $N$-vortex states of a 2D BEC and the microstates of
the classical point-vortex model.  The set of all microstates $\psi$ at fixed
point-vortex energy $\varepsilon$ \eqnrefp{eqn:pv_energy} defines a
microcanonical ensemble; the measure of this set is the structure function
$W(\varepsilon)$ [which defines the system entropy $S(\varepsilon) = k_B
\ln(W(\varepsilon))$].  The normalized structure function, $w(\varepsilon)\equiv W(\varepsilon)(\int d\varepsilon W(\varepsilon))^{-1}$, is obtained
numerically as a histogram of $\varepsilon$ for random vortex
configurations.  We sample the microcanonical ensemble at energy $\varepsilon$
numerically, using a random walk to generate many $N$-vortex configurations
having energies within a given tolerance \cite{sampling}. Related
microcanonical sampling techniques have previously been applied to the classical
point-vortex model \cite{campbell_oneil_jsp_1991, weiss_mcwilliams_pf_1991,
yatsuyanagi_etal_prl_2005, sano_etal_jpsj_2007}.  Averages of observables over
this ensemble are dominated by the most likely (highest-entropy)
configurations.  For large $N$ (ensuring ergodicity
\cite{weiss_mcwilliams_pf_1991, campbell_oneil_jsp_1991}) ensemble
averages define a \textit{statistical equilibrium} corresponding to
time-averaged properties of the end-states of decaying quantum vortex turbulence
at energy $\varepsilon$. 

To demonstrate that quantum vortices in a 2D BEC can provide a
physical realization of negative-temperature states exhibiting macroscopic
vortex clustering, we sample the GPE microstates
of the 2D BEC, compute the IKE spectrum, and decompose the vortex configurations into dipoles and clusters using the
recursive cluster algorithm (RCA) developed in
Ref.~\cite{reeves_etal_prl_2013}. For each cluster the RCA yields the cluster
charge $\kappa_c$ and average radius $r_c$ (average distance of constituent vortices
from the cluster center of mass). We define the clustered fraction $f_c =
\sum_s |\kappa_{c,s}| / N$, where $\kappa_{c,s}$ is the charge of the $s$th
cluster. We also define $N_{\kappa_c}$ as the total number of vortices participating
in all clusters of charge $\pm |\kappa_c|$.  Finally, we introduce the correlation
functions $c_B = \sum_{p=1}^N \sum_{q=1}^B \kappa_p \kappa^{(q)}_p / BN$, where
$\kappa^{(q)}_p$ is the charge of the $q$th nearest-neighbor to vortex $p$.
These are directly related to the functions $C_B$ introduced in
Ref.~\cite{White2012a}; a value of $c_B > 0$ ($<0$) indicates
(anti-)correlation between vortex charges, up to nearest neighbours of $B$th
order.

\begin{figure*}
\includegraphics[width=0.95\textwidth]{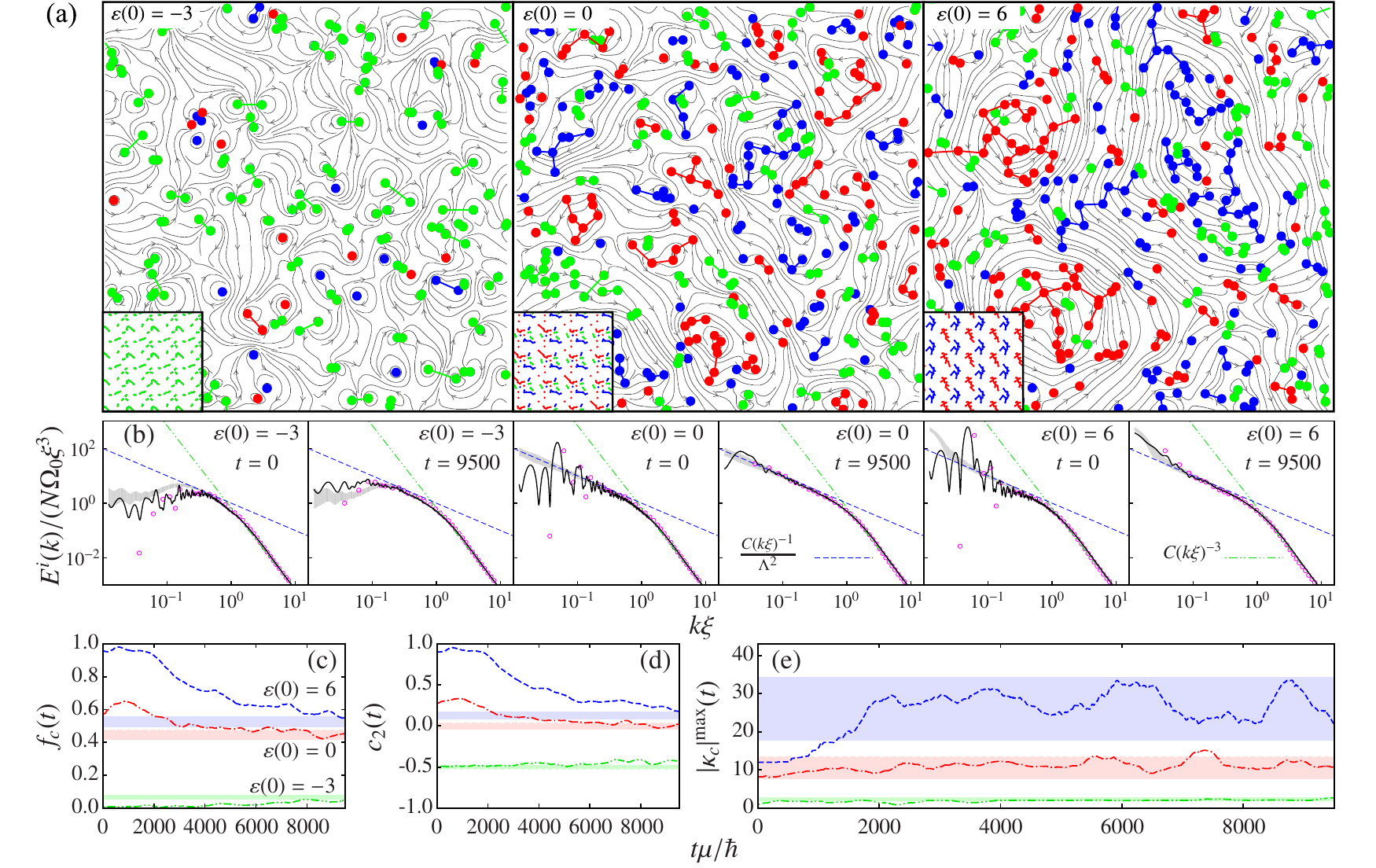}
\caption{(color online). Dynamical dGPE evolution of non-equilibrium neutral $N$-vortex
states towards statistical equilibrium (see also animations in \cite{suppinf}).
(a): RCA decomposition after equilibriation ($t=9500\hbar/\mu$), with initial
conditions inset. Streamlines show the incompressible velocity field. Field of
view is $L\times L$. (b): IKE spectrum, compared to the statistical equilibrium
distribution from \figreft{figure_Sampling}{(c)} (grey solid line, thickness
indicates $\pm1$ standard deviations). Other symbols in (a,b) as in
\figreft{figure_Sampling}{(c)}. (c): Clustered fraction $f_c$. (d): Correlation
function $c_2$.  (e): Absolute charge, $|\kappa_c|$, of the largest vortex
cluster. Spectra and measures are shown as a moving average from $t$ to
$t+500\hbar/\mu$, and $E^i(k)$ is normalized by $N(t)$; decay of $N(t)$ for
$\varepsilon(0)>-3$ is negligible ($\lesssim 5\%$).  Horizontal shading in
(c--e) shows the statistical equilibrium distributions from
\figreft{figure_Sampling}{(a,b)} (shaded areas indicate $\pm1$ standard
deviations). Note that we compare the $\varepsilon(0)=-3$ evolution to the
statistical equilibrium at $\varepsilon=-2.5$, since dipole annihilation [for
$\varepsilon=-3$, $N(10^4) = 178$] leads to $\varepsilon(10^4) \approx -2.5$.
\label{figure_TimeEvolution}}
\end{figure*}

\figreftfull{figure_Sampling}{(a)} shows $w(\varepsilon)$ for $N=384$ vortices
in a doubly periodic box of side $L=512\xi$.  The boundary between positive- and
negative-temperature states, $\varepsilon_\infty$, lies at the maximum of
$W(\varepsilon)$, where the temperature $T= W (\partial W/\partial
\varepsilon)^{-1}/k_B \rightarrow \infty$. We find $\varepsilon_\infty
\approx -0.255$ and the mean energy $\langle \varepsilon \rangle = 0$ known from
the point-vortex model \cite{campbell_oneil_jsp_1991} despite the
minimum-separation constraint.  \figreftfull{figure_Sampling}{} also shows the
averages of the clustering measures
\figrefp{figure_Sampling}{(a,b)}, the IKE spectrum
\figrefp{figure_Sampling}{(c)}, and the distribution of $N_{\kappa_c}$
\figrefp{figure_Sampling}{(d)} as a function of $\varepsilon$.  At $\varepsilon
= 0$, $c_B$ ($f_c$) is equal to $0$ ($1/2$), indicating an
uncorrelated vortex distribution. The distribution of $N_{\kappa_c}$ is
strongly skewed towards small clusters, with $|\kappa_c|^{\rm max} < 20$ and
$r_c^{\rm max} < 50\xi\ll L$, and the
IKE spectrum follows the $k^{-1}$ law in the IR-region.  At energies
$\varepsilon < 0$ (where $T>0$ for $\varepsilon \le \varepsilon_\infty$), $c_B$
($f_c$) drops below $0$ ($1/2$), indicating proliferation of vortex dipoles and reduced number, charge, and radius of vortex clusters. As
$\varepsilon \rightarrow -3$ one obtains a vortex-dipole gas with approximately
the minimal spacing $2\pi \xi$ [see \figreft{figure_TimeEvolution}{(a)}]. The
IKE spectrum lies below the $k^{-1}$ law at large scales.  At energies
$\varepsilon > 0$ (where $T<0$) \figreft{figure_Sampling}{} shows macroscopic vortex clustering and spectral
condensation of IKE. While low-order measures of clustering
($c_B$, $f_c$) increase slowly with $\varepsilon$, the distribution of vortices bifurcates, revealing the appearance of two
(opposite-sign) macroscopic clusters. Spectrally, the energy associated with
these clusters manifests itself as an OKC lying above the $k^{-1}$ power-law at
large scales. The charge and radius of clusters ($|\kappa_c|^{\rm max}$
and $r_c^{\rm max}$) grows more rapidly with $\varepsilon$ than $\langle
|\kappa_c| \rangle$ and $\langle r_c \rangle$ up to $\varepsilon \sim 50$,
highlighting the utility of the RCA for the characterization of point-vortex
states. Above this energy, clusters increase in charge less
rapidly, and absorb further energy by shrinking in radius. For $\varepsilon
\gtrsim 200$ two clusters contain all the vortices, illustrating
the phenomenon of supercondensation \cite{Kra1980.RPP5.547}; in this regime the
$k^{-1}$ spectrum vanishes.

In contrast to the UV-divergent point-vortex model, the universal $k^{-3}$
 UV asymptotic of the IKE spectrum \eqnrefp{eqn:2dqt_spectrum} implies a
\textit{physical} transition energy for the emergence of the OKC, given by $E_0
\equiv E^i_{\rm tot}(\varepsilon=0)$, where $E^i_{\rm tot}(\varepsilon)\equiv\int dk E^i(k)$. An analytic estimate of $E_0$ follows
from the second term in \eqnreft{eqn:2dqt_spectrum} averaging to zero at
$\varepsilon=0$ ($c_B=0$ for all $B$); correctly accounting for the
discrete nature of the spectrum \cite{suppinf} yields $E_0  \approx 4.735
N\Omega_0 \xi^2$, in good agreement with the numerical value $E_0
\approx 4.821 N \Omega_0 \xi^2$ obtained from $\psi$, which does not rely on
the core ansatz used to obtain \eqnreft{eqn:2dqt_spectrum}. We find that the
total IKE is very well predicted by $E^i_{\rm
tot}(\varepsilon) = E_0 + \varepsilon N \Omega_0 \xi^2$.  Thus, the appearance
of the OKC is due to saturation of excited states: IKE
exceeding $E_0$ accumulates near the system scale, forming the spectral
feature lying above the $k^{-1}$ spectrum.  Note that the OKC is a (quasi-)equilibrium phenomenon distinct from the 
classical scenario of condensation in \textit{forced} turbulence, where the
$k^{-5/3}$ spectrum of the inverse energy cascade (IEC) gives way to $k^{-3}$ at low
$k$ associated with the dynamical condensate
\cite{Chertkov2007a,xia_etal_prl_2008,chan_etal_pre_2012}.  

To demonstrate that \figreft{figure_Sampling}{} provides a quantitative
description of decaying QT in a 2D BEC, and that statistically-driven transfers
of energy to large length scales can occur in a \textit{compressible} quantum
fluid, we consider the dynamics of non-equilibrium states in the
dGPE~\cite{Tsubota2002,Penckwitt2002,Blakie08a}.  For a 2D BEC (subject to
tight harmonic confinement in the $z$-direction with oscillator length $l_z$)
this can be written as 
\begin{equation}
i\hbar \frac{\partial \psi
(\mathbf{r},t)}{\partial t} = (1-i\gamma) \left(
-\frac{\hbar^2\nabla_\perp^2}{2m} + g_2|\psi(\mathbf{r},t)|^2 - \mu \right)
\psi(\mathbf{r},t)\,, 
\end{equation} 
where $g_2 = \sqrt{8\pi}\hbar^2 a_s/ml_z$, $m$ is the atomic mass, and $a_s$ is
the $s$-wave scattering length. The dimensionless damping rate $\gamma$
describes collisions between condensate atoms and non-condensate atoms, an
important physical process in real 2D superfluids that leads to effective
viscosity \cite{Bradley2012a} and suppression of sound energy at high $k$.  We
use the experimentally realistic value $\gamma=10^{-4}$ \cite{neely_etal_prl_2013} in
our simulations.  We use a random walk to obtain a neutral
configuration of $N'=24$ vortices in a periodic box of length $L'=128\xi$ at
energy $\varepsilon'$. A $4^2$ tiling of this configuration, with each vortex
subject to Gaussian position noise (variance $\xi$), provides a
non-equilibrium, low-entropy state of $N=384$ vortices in a box with $L=512\xi$
and energy $\varepsilon$ (found by adjusting $\varepsilon^\prime$).  This state
is then evolved to time $10^4\hbar/\mu$ in the dGPE, over which time we find that
the compressible energy does not increase [and $\varepsilon(t)$ does not decay]
significantly, supporting our statistical description \cite{gzero}.

\figreftfull{figure_TimeEvolution}{} shows the time evolution of the vortex
configuration for three different initial
energies~\figrefp{figure_TimeEvolution}{(a)} with IKE spectra
\figrefp{figure_TimeEvolution}{(b)} and clustering measures
\figrefp{figure_TimeEvolution}{(c-e)} determined by short-time averaging of
individual runs of the dGPE. While approach to complete equilibrium is slow,
\figreft{figure_TimeEvolution}{(e)} shows that the charge of the largest
cluster equilibriates more rapidly (by $t \sim 2000 \hbar /\mu$). For
$\varepsilon(0)=-3$, the dynamics consists largely of dipole-dipole collisions
and vortex-antivortex annihilation (\textit{increasing} the energy per vortex
$\varepsilon$, \cite{campbell_oneil_jsp_1991}), and exhibits a time-invariant IKE spectrum; these
positive-temperature point-vortex states have no analog in
2DCT~\cite{campbell_oneil_jsp_1991}. For $\varepsilon(0)=0$, the approach to
equilibrium involves significant dipole-cluster interactions that redistribute
the cluster charges, decreasing $f_c$ and $c_B$ while increasing
$|\kappa_c|^{\rm max}$.  This redistribution transfers energy to large scales,
producing an approximate $k^{-1}$ power-law in the IR [note that the
time-averaged spectrum is expected to fluctuate relative to the
$10^4$-configuration average in Fig. 1(c)]. For $\varepsilon(0)=6$ the dynamics
is reminiscent of 2DCT.  Energy transfer to large scales builds an OKC, with
the vortices grouping into two macroscopic clusters \cite{okcV}.  Although a
steady $k^{-5/3}$ spectrum is absent (and would require continuous forcing and
damping to establish a steady inertial range \cite{reeves_etal_prl_2013}) some
intermittent $k^{-5/3}$ behaviour is evident \cite{suppinf}.  The equilibrium
distribution of vortices outside of the OKC closely resembles the uncorrelated
($\varepsilon=0$) state \cite{Eyi2006.RMP78.87} and has a low level of
clustering. As the initial condition contains many small clusters, this
counterintuitively causes low-order measures of clustering [$f_c$ and $c_2$ in
\figreft{figure_TimeEvolution}{(c,d)}, and all $c_B$ for $B \lesssim 10$] to
\textit{decay} during OKC formation. Thus, high-order clustering information
provided by the RCA is vital in identifying OKC: the rapid increase of
$|\kappa_c|^{\rm max}$ for $\varepsilon(0)=6$ in
\figreft{figure_TimeEvolution}{(e)} contrasts with the cases
$\varepsilon(0)=0,-3$, indicating the emergence of the OKC. The demonstration
of a statistically driven transfer of kinetic energy to large scales underpins
the existence of an IEC in far-from-equilibrium 2DQT in scenarios with
appreciable vortex clustering~\cite{Bradley2012a,reeves_etal_prl_2013}, and is
complementary to the direct energy cascade identified in scenarios dominated by
vortex-dipole recombination~\cite{Numasato10b,CheslerScience2013}. 

We have developed a first-principles realization of Onsager's point-vortex
model in a 2D superfluid, and observe the upscale energy transfer of 2DCT in decaying 2DQT described by the damped Gross-Pitaevskii equation. Configurational analysis of the vortex states and associated energy spectra demonstrate the emergence of an Onsager--Kraichnan condensate of
quantum vortices occurring at negative temperatures in equilibrium, and as the end states of decaying 2DQT. The microcanonical sampling approach
opens a new direction in the study of 2DQT, enabling systematic studies of far-from-equilibrium dynamics,
energy transport, inertial ranges, and other emergent phenomena in 2DQT, and points the way to experimental realization of Onsager--Kraichnan
condensation.
\acknowledgments
We thank P. B. Blakie and A. L. Fetter
for valuable comments. This work was supported by The New Zealand Marsden Fund, and a Rutherford Discovery Fellowship of the
Royal Society of New Zealand. BPA is supported by the US National
Science Foundation (PHY-1205713).  We are grateful for the use of NZ eScience
Infrastructure HPC facilities (http://www.nesi.org.nz).

%

\section*{Supplemental Material}
\setcounter{equation}{0}
\setcounter{figure}{0}

\section{I. Construction of the quantum phase of an $N$-vortex wavefunction in a periodic square domain}
Here, we construct the phase $\theta(\rr,\{\rr_j\},\{\kappa_j\})$ associated
with an ansatz wavefunction for a neutral system of $N$ superfluid vortices
with positions $\rr_j$ and circulations $h \kappa_j / m$ (charges $\kappa_j
= \pm1$) in a homogeneous periodic square 2D BEC of side $L$.

The phase associated with the velocity field of a single point-vortex $j$
is given by 
\begin{equation}
\theta_j = \kappa_j \mbox{atan2}(y-y_j,x-x_j)\,,
\label{eqn:full_angle}
\end{equation}
where the four-quadrant arctangent function is defined by
\begin{equation}
\mbox{atan2} (y,x) = \left\{ 
\begin{array}{lr} 
\atan(y/x), & x>0 \\ 
\atan(y/x)+\pi, & x<0,y>0 \\
\atan(y/x)-\pi, & x<0,y<0 
\end{array} \right.
\end{equation}
Because we only require $\theta_j$ to within an arbitrary multiple of $2\pi$,
we may gain a substantial notational convenience by noting that the simplified
definition
\begin{equation}
\theta_j = \kappa_j \atan(y-y_j,x-x_j) + \pi H(x-x_j)\,,
\label{eqn:angle_simple}
\end{equation}
where $H(x)$ is the Heaviside unit-step function, is equivalent to
\eqnreft{eqn:full_angle} to within such a multiple.  Due to the infinite extent
of this phase field, to obtain the phase due to $N$ point-vortices in a
doubly-periodic square domain it is necessary to sum over the entire infinite
periodic array of vortices (i.e., not only over the periodic unit cell, but
over the infinite periodic lattice).  Implemented directly, such a summation is
poorly convergent. Summing over too few unit cells of the infinite lattice
introduces spurious boundary effects at the edge of the central unit cell; in
particular the nucleation of undesired extra vortices at the boundaries leading
to subsequent large-scale compressible flows.  While this problem can be to
some extent mitigated by a short evolution in imaginary time, by performing
such an evolution one relinquishes control over the exact number and positions
of the vortices.  However, when computing the phase on a large simulation grid,
even for small numbers of vortices ($< 10$), we find that summing over a
sufficient number of unit cells to eliminate spurious boundary effects is
computationally infeasible.  In the following we overcome this challenge by
analytically reducing the poorly-convergent, doubly-infinite sum to a single
convergent summation. The final form for $\theta$ we present allows exact
vortex positioning, eliminates boundary effects entirely, and leads to very
short computation times even on large grids.

\begin{figure}
\includegraphics[width=\columnwidth]{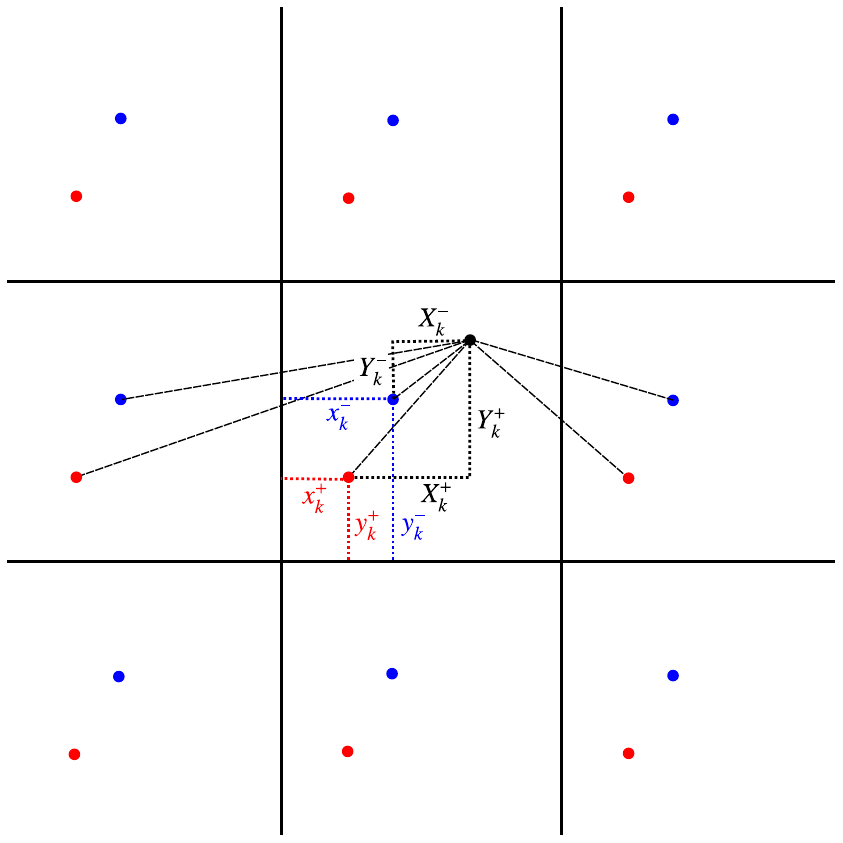}
\caption{Diagram illustrating summation over the infinite periodic array of
vortices corresponding to a single vortex dipole in the unit cell. The black
dot indicates a test position $(x,y)$ at which we compute the phase due to the
dipole formed by a positive-circulation vortex at $(x_k^+,y_k^+)$ and a
negative-circulation vortex at $(x_k^-,y_k^-)$.  }
\label{fig:drawing}
\end{figure}

We begin by working in a periodic unit cell with coordinates $0\le x,y<2\pi$
for simplicity (the extension to the general periodic cell $0 \le
\tilde{x},\tilde{y}<L$ is straightforward and follows subsequently).
\figreftfull{fig:drawing}{} illustrates the geometry of the problem for the
$k$th vortex dipole (any neutral configuration of $N$ vortices can be
arbitrarily partitioned into $N/2$ dipoles).  This vortex dipole is formed by a
positive vortex located at $(x_k^+,y_k^+)$ and a negative vortex located at
$(x_k^-,y_{k}^-)$.  To determine the phase at a test point $(x,y)$ it is
convenient to introduce the auxilliary variables $(X_k^+,Y_k^+) =
(x-x_k^+,y-y_k^+)$ and $(X_k^-,Y_k^-)=(x-x_k^-,y-y_k^-)$. In the present
treatment we shall assume that $|x_k^+-x_k^-|,|y_k^+-y_k^-|<\pi$; this
restriction can always be met by an appropriate translation of the unit cell
and provides a considerable notational convenience. However, note that when
computing the phase due to a given $N$-vortex configuration it is in fact
somewhat easier to follow the procedure developed here for each vortex pair
\textit{without} enforcing this requirement: Instead one can subsequently
transform the phase $\theta_k$ for each vortex dipole according to $\theta_k
\rightarrow \theta_k - x H(|y_k^+-y_k^-|-\pi) + y H(|x_k^+-x_k^-|-\pi)$ to
obtain the correct result.

We proceed by first reducing the doubly-infinite summation to a singly-infinite
one by analytically summing over all periodic replicas of the unit cell in the
$x$-direction.  Explicitly considering just the central row of unit cells in
the figure, we have for periodic replicas of the positive vortex
\begin{multline}
  \theta_{k,+}^{(\mathrm{row})}(x,y) = \sum_{m=0}^\infty \atan\left( \frac{Y_k^+}{2\pi m + X_k^+} \right) - \sum_{m=1}^\infty \atan \left(\frac{Y_k^+}{2\pi m - X_k^+} \right) \\+ \pi H(X_k^+) + f^+(y)\,.
\label{eqn:basic_sum}
\end{multline}
The penultimate term above represents the adjustment to the angle obtained when ``passing'' a vortex in the $x$-direction [see \eqnreft{eqn:angle_simple}]. The final term reflects the fact that the phase is only defined up to a, possibly $y$-dependent, constant. Rearranging the sums in \eqnreft{eqn:basic_sum} under a single summation sign we obtain
\begin{multline}
\theta_{k,+}^{(\mathrm{row})}(x,y) = \sum_{m=1}^\infty \left[ \atan\left( \frac{Y_k^+}{2\pi m + (X_k^+-\pi) - \pi} \right) \right.\\ \left. - \atan \left( \frac{Y_k^+}{2\pi m - (X_k^+-\pi) - \pi} \right) \right] + \pi H(X_k^+) +f^+(y) \,.
\end{multline}
This summation can be evaluated using a formula due to Ramanujan \cite{ramanujan_sum}, giving
\begin{multline}
  \theta_{k,+}^{(\mathrm{row})}(x,y) = -\atan \left[ \tanh \left( \frac{Y_k^+}{2} \right) \tan \left( \frac{X_k^+-\pi}{2} \right) \right] \\+ \pi H(X_k^+) + f^+(y)\,.
\end{multline}
An equivalent treatment for the negative vortex yields
\begin{multline}
  \theta_{k,-}^{(\mathrm{row})}(x,y) = \atan \left[ \tanh \left( \frac{Y_k^-}{2} \right) \tan \left( \frac{X_k^--\pi}{2} \right) \right] \\- \pi H(X_k^-) + f^-(y)\,.
\end{multline}

Summing over all the rows of unit cells (i.e., summing over the $y$-direction) leads to the singly-infinite sum for the total phase due to the dipole and all its periodic replicas
\begin{multline}
\theta_k(x,y) = \sum_{n=-\infty}^\infty \left\{ \atan \left[\tanh \left( \frac{Y_k^- + 2\pi n}{2} \right) \tan \left(\frac{X_k^- - \pi}{2} \right) \right] \right. \\ \left.- \atan \left[\tanh \left(\frac{Y_k^+ + 2\pi n}{2} \right) \tan \left(\frac{X_k^+ - \pi}{2} \right) \right]   \right. \\
\left. +\pi \left[H(X_k^+) - H(X_k^-) \right] \vphantom{\tanh \left( \frac{Y_k^- + 2\pi n}{2} \right)} \right\} + f(y)\,,
\label{eqn:big_sum_expr}
\end{multline}
where $f(y) = f^+(y)+f^-(y)$. The correct $f(y)$ is determined by the requirement that the phase be periodic in $y$, in the sense that 
\begin{equation}
\label{eqn:phase_restriction}
\theta_k(x,y+2\pi) = \theta_k(x,y) + s2\pi\,,
\end{equation}
for integer $s$.
Keeping in mind this requirement, we differentiate \eqnreft{eqn:big_sum_expr} w.r.t. $y$, yielding the $y$-component of the velocity field
\begin{multline}
{v_{y}}_k(x,y) = \frac{1}{2} \sum_{n=-\infty}^\infty \left[\frac{-\sin(X_k^+)}{\cos(X_k^+)-\cosh(2 n \pi +Y_k^+)}\right. \\\left.+\frac{\sin(X_k^-)}{\cos(X_k^-)-\cosh(2n \pi +Y_k^-)}\right] + f'(y)\,.
\label{eqn:full_derivative}
\end{multline}
Note that this differs from the standard point-vortex result \cite{weiss_mcwilliams_pf_1991} only by the term $f'(y)$. To obtain appropriate point-vortex physics, we therefore require that $f'(y)$ be equal to a constant. Physically, this constant appears because of the 
restriction \eqnrefp{eqn:phase_restriction} that $\theta_k$ should be a well-defined quantum phase, a requirement not present in the classical point-vortex model. \eqnreftfull{eqn:full_derivative} can be integrated over a single period in $y$ to obtain the change in phase over the unit cell in this direction,
\begin{multline}
\Delta \theta_k = \int_0^{2\pi} dy {v_{y}}_k(x,y) \\= \frac{1}{2} \int_{-\infty}^\infty dy \left[\frac{-\sin(X_k^+)}{\cos(X_k^+)-\cosh(Y_k^+)}\right. \\\left.+\frac{\sin(X_k^-)}{\cos(X_k^-)-\cosh(Y_k^-)}\right] + \int_0^{2\pi} dy f'(y)\,,
\end{multline}
where we have used the periodicity of ${v_{y}}_k$ in order to replace the infinite sum in \eqnreft{eqn:full_derivative} with infinite limits in the integral.

Evaluating both integrals, for constant $f'(y)$, gives
\begin{equation}
\Delta \theta_k = x_k^+ - x_k^- + 2\pi f'(y)\,.
\end{equation}
Hence, the minimal choice to recover periodicity in $y$ is $f'(y)=-(x_k^+-x_k^-)/2\pi$, yielding the final expression
\begin{multline}
\theta_k(x,y) = \sum_{n=-\infty}^\infty \left\{ \atan \left[\tanh \left( \frac{Y_k^- + 2\pi n}{2} \right) \tan \left(\frac{X_k^- - \pi}{2} \right) \right] \right. \\ \left.- \atan \left[\tanh \left(\frac{Y_k^+ + 2\pi n}{2} \right) \tan \left(\frac{X_k^+ - \pi}{2} \right) \right]   \right. \\
\left. +\pi \left[H(X_k^+) - H(X_k^-) \right] \vphantom{\tanh \left( \frac{Y_k^- + 2\pi n}{2} \right)} \right\} - \frac{x_k^+-x_k^-}{2\pi} y\,.
\label{eqn:final_sum_expr}
\end{multline}
In practice the sum in \eqnreft{eqn:final_sum_expr} converges rapidly, such that $\sum_{n=-\infty}^{\infty} \rightarrow \sum_{n=-5}^{5}$ is sufficient.

Using the dipole result above for the case $L=2\pi$, the total phase due to a neutral configuration of $N$ vortices in a periodic box of arbitrary side length $L$, with coordinate system $(\tilde{x},\tilde{y})\equiv \tilde{\rr}$, can be obtained by an appropriately scaled sum over all dipoles
\begin{equation}
\theta_{\rm total}(\tilde{x},\tilde{y}) = \sum_{k=1}^{N/2} \theta_k(x,y)\,,
\label{eqn:phase_ansatz}
\end{equation}
where $(x,y) = 2\pi (\tilde{x},\tilde{y})/L$, and the implicit vortex position arguments to $\theta_k$ should also be appropriately scaled (i.e., $\rr_j = 2\pi \tilde{\rr}_j /L$).
We have tested this expression for a wide variety of numbers, $N$, and individual configurations of vortices. We find that the result is periodic with period $L$ and, importantly, the velocity field $\hbar \nabla \theta_{\rm total} /m$ associated with the phase agrees exactly with the point-vortex result of Ref.~\cite{weiss_mcwilliams_pf_1991}, up to the small velocity shift required to obtain a well-defined quantum phase. 

Finally, using \eqnreft{eqn:final_sum_expr} and \eqnreft{eqn:phase_ansatz} we construct the GPE ansatz state for a neutral configuration of $N$ vortices in a periodic box of arbitrary side length $L$
\begin{equation}
\psi(\tilde{x},\tilde{y}) = e^{i \theta_{\rm total}(\tilde{x},\tilde{y})} \prod_{p=1}^N \chi \left( \sqrt{(\tilde{x}-\tilde{x}_p)^2+(\tilde{y}-\tilde{y}_p)^2} \right)\,,
\end{equation}
where $\chi(\tilde{r})$ is the radial profile of an isolated quantum vortex core \cite{Bradley2012a}, which we obtain numerically. To ensure the cores are indeed suitably isolated, we enforce a minimum initial separation between vortices of $2\pi \xi$, where $\xi$ is the healing length. Evolving this initial condition in the undamped GPE, with no preparatory imaginary-time evolution or other smoothing, leads to negligible density fluctuations during the first part of the ensuing vortex dynamics, and no unphysical dynamics at the boundary of the periodic cell. Hence, \eqnreft{eqn:phase_ansatz} provides a suitable ansatz for an arbitrary $N$-vortex state (with minimum inter-vortex distance $2 \pi \xi$) in the homogeneous, periodic GPE.

\section{II. Point-vortex energy spectrum in a periodic square domain}
For $N$ vortices at positions $\rr_i$ with charges $\kappa_i$ in a periodic square box of side $L\xi$, the stream function $\varphi(\rr)$ obeys
\begin{equation}
-\nabla^2 \varphi(\rr;\{\rr_i,\kappa_i\}) = 2\pi \xi c \sum_{i=1}^N \kappa_i \delta^{(2)} (\rr-\rr_i),
\end{equation}
where $2\pi \xi c = h/m$ is the quantum of circulation and $\delta^{(2)}(\rr)$ is the two-dimensional Dirac delta.

Green's functions provide a powerful method for obtaining the stream function for superfluid vortices \cite{fetter_jltp_1974}.
The stream function can be derived as a sum of single-vortex Green's functions which obey
\begin{equation}
-\nabla^2 G(\rr,\rr_i,\kappa_i) = \kappa_i \delta^{(2)}(\rr-\rr_i).
\label{eqn:green}
\end{equation}
\eqnreftfull{eqn:green} has general solution under our periodic boundary conditions
\begin{equation}
G(\rr,\rr_i,\kappa_i) = \kappa_i \sum_{\kk \ne 0} \frac{e^{i\kk\cdot (\rr-\rr_i)}}{L^2 \xi^2 k^2},
\end{equation}
where $\kk = 2\pi(n_x,n_y)/L\xi$ for $n_x,n_y \in \mathbb{Z}$.
Hence, the stream function $\varphi$ is given by
\begin{equation}
\varphi(\rr;\{\rr_i,\kappa_i\}) = \sum_{i=1}^N \sum_{\kk \ne 0} 2\pi \xi c \kappa_i \frac{e^{i\kk\cdot (\rr-\rr_i)}}{L^2 \xi^2 k^2}.
\end{equation}
The two dimensional velocity field $\vvv = \nabla \times \varphi \hat{\mathbf{z}}$ is thus
\begin{equation}
\vvv(\rr;\{\rr_i,\kappa_i\}) = \sum_{i=1}^N \sum_{\kk \ne 0} \kappa_i \frac{2\pi c}{L^2 \xi} \frac{ie^{i\kk\cdot (\rr-\rr_i)}}{k^2} 
\left( \begin{array}{c} k_y \\ -k_x \end{array} \right).
\label{eqn:vel}
\end{equation}
Fourier transforming \eqnreft{eqn:vel} to obtain
\begin{equation}
\ww(\kk;\{\rr_i,\kappa_i\}) = \frac{1}{2\pi} \int\limits_0^{L\xi} \int\limits_0^{L\xi} d\rr\; \vvv(\rr;\{\rr_i,\kappa_i\}) e^{-i\kk \cdot \rr}, 
\end{equation}
yields
\begin{equation}
\ww(\kk;\{\rr_i,\kappa_i\}) = \sum_{i=1}^N  \kappa_i c \xi \frac{ie^{-i\kk\cdot \rr_i}}{k^2} 
\left( \begin{array}{c} k_y \\ -k_x \end{array} \right).
\label{eqn:ftvel}
\end{equation}
Hence the kinetic energy spectrum $E(\kk) = m n_0 |\ww(\kk)|^2/2$, for background superfluid number density $n_0$, is given by
\begin{align}
E(\kk) &=   \frac{m n_0 c^2 \xi^2}{2k^2} \left|\sum_{i=1}^N \kappa_i e^{-i\kk\cdot \rr_i}\right|^2, \nonumber\\
 &=  \frac{m n_0 c^2 \xi^2}{2k^2} \left[ N + 2\sum_{i=1}^{N-1} \sum_{j=i+1}^N \kappa_i \kappa_j \cos[\kk\cdot (\rr_i-\rr_j)] \right],\nonumber\\
 &=  \frac{N \Omega_0 \xi^4}{2\pi (k\xi)^2} \left[ 1 + \frac{2}{N}\sum_{i=1}^{N-1} \sum_{j=i+1}^N \kappa_i \kappa_j \cos[\kk\cdot (\rr_i-\rr_j)] \right],
\label{eqn:2dspec}
\end{align}
where $\Omega_0 = \pi \hbar^2 n_0/m\xi^2$ is the quantum of enstrophy \cite{Bradley2012a}.

While \eqnreft{eqn:2dspec} gives the full form of the kinetic energy spectrum, it is typically more practical to consider the angularly integrated spectrum $E(k)$. By assuming a continuum limit and substituting $k_x = k\cos(\theta_k)$, $k_y = k\sin(\theta_k)$, and $x_i-x_j = r_{ij}\cos(\theta_{ij})$, $y_i-y_j = r_{ij} \sin(\theta_{ij})$, one can obtain the angularly integrated spectrum
\begin{align}
E(k) &= k\int_0^{2\pi} d\theta_k\; E(\kk),\nonumber \\
&= \frac{N\Omega_0 \xi^3}{k\xi} \left[1 + \frac{2}{N} \sum_{i=1}^{N-1} \sum_{j=i+1}^N \kappa_i \kappa_j J_0 (kr_{ij}) \right].
\end{align}
To obtain an incompressible kinetic energy with the correct ultraviolet
asymptotics for vortices in a compressible quantum fluid, one can follow the
procedure of Ref.~\cite{Bradley2012a} and use an ansatz for the vortex core density.
This procedure leads to an overall envelope function on the spectrum which is
exactly equivalent to the replacement $(k\xi)^{-1} \rightarrow F_\Lambda
(k\xi)$. Hence
\begin{equation}
E^i(k) =  N\Omega_0 \xi^3 F_\Lambda(k\xi) \left[1 + \frac{2}{N} \sum_{i=1}^{N-1} \sum_{j=i+1}^N \kappa_i \kappa_j J_0 (kr_{ij}) \right],
\label{eqn:final_spectrum}
\end{equation}
is the appropriate kinetic spectrum for $N$ superfluid vortices in a periodic square box, in the Gross-Pitaevskii description.

\section{III. Estimates of the Onsager--Kraichnan condensation energy}
Since, unlike the point-vortex model, the Gross-Pitaevskii description of superfluid vortices is UV-convergent, it is interesting to consider the predicted transition energy, $E_0$, at which the Onsager--Kraichnan condensate emerges. At the transition point, $\varepsilon=0$ , vortices are uncorrelated and the Bessel part of the spectrum Eq. (\ref{eqn:final_spectrum}) averages to zero, leaving $N$ times the single-vortex spectrum. That is,
\begin{align}
E^i_{\varepsilon=0}(k) = N \Omega_0 \xi^3 F_\Lambda(k\xi)\,.
\end{align}
Integrating up to the largest available scale, one obtains the estimate for the transition energy $E_0 = E^i_{\rm tot}(\varepsilon=0)$:
\begin{equation}
E_0 = \int_{2\pi/L}^\infty E^i(k) dk \approx 4.324 N \Omega_0 \xi^2\,.
\end{equation}
However, this estimate neglects the discrete nature of the modes around $k=2\pi/L$, and hence does not give a particularly accurate value for $E_0$.
A better estimate is obtained by applying the core ansatz to renormalize the spectrum as a function of $\mathbf{k}$:
\begin{align}
E^i(\mathbf{k}) = \frac{N \Omega_0 \xi^4 F_\Lambda(k\xi)}{2\pi k\xi} \left[ 1 + \frac{2}{N} \sum_{i=1}^{N-1} \sum_{j=i+1}^N \kappa_i \kappa_j \cos(\mathbf{k}\cdot\mathbf{r}_{ij}) \right]\,;
\end{align}
similarly neglecting the cosine terms and summing over the discrete modes $\mathbf{k}=(2\pi n_x/L,2\pi n_y /L)$ yields $E_0 \approx 4.735 N \Omega_0 \xi^2$. This estimate compares well with the average value obtained directly from the uncorrelated $N$-vortex ansatz state $\psi$ 
(which uses the correct numerical vortex core solution, rather than the approximate ansatz implicit in $F_\Lambda$) of $E_0 \approx 4.821 N \Omega_0 \xi^2$.

\section{IV. Dynamical evolution and role of damping}

\subsection{A. Damped Gross--Pitaevskii model}
We model the dynamics of a compressible
two-dimensional BEC within the framework of the damped Gross-Pitaevskii
equation (dGPE)~\cite{Tsubota2002,Penckwitt2002,Blakie08a} 
\begin{equation} 
i\hbar \frac{\partial \psi (\mathbf{r},t)}{\partial t} = (1-i\gamma) \left(
-\frac{\hbar^2\nabla_\perp^2}{2m} + g_2|\psi(\mathbf{r},t)|^2 - \mu \right)
\psi(\mathbf{r},t)\,, 
\end{equation}
where $g_2 = \sqrt{8\pi}\hbar^2 a_s/ml_z$, $m$ is the atomic mass, $a_s$ is the
$s$-wave scattering length, $\mu$ is the chemical potential, and $l_z$ is the
oscillator length associated with the (tightly confining) external harmonic trap
in the $z$-direction. 
As stated in our letter, the dimensionless damping rate $\gamma$ describes
collisions between condensate atoms and non-condensate atoms with
chemical potential $\mu$.  These collisions are an important physical process
in real 2D superfluids. A wide-ranging theoretical framework for dealing with these effects at
different levels of approximation is provided by $c$-field theory
\cite{Blakie08a}. Within this framework, the dGPE is obtained as a
low-temperature approximation to the simple growth form
\cite{bradley_etal_pra_2008} of the stochastic projected Gross-Pitaevskii
equation (SPGPE) \cite{gardiner_davis_jpb_2003} by neglecting thermal noise,
but retaining dissipation in the form of the rate $\gamma$. 

\begin{figure}
\includegraphics[width=\columnwidth]{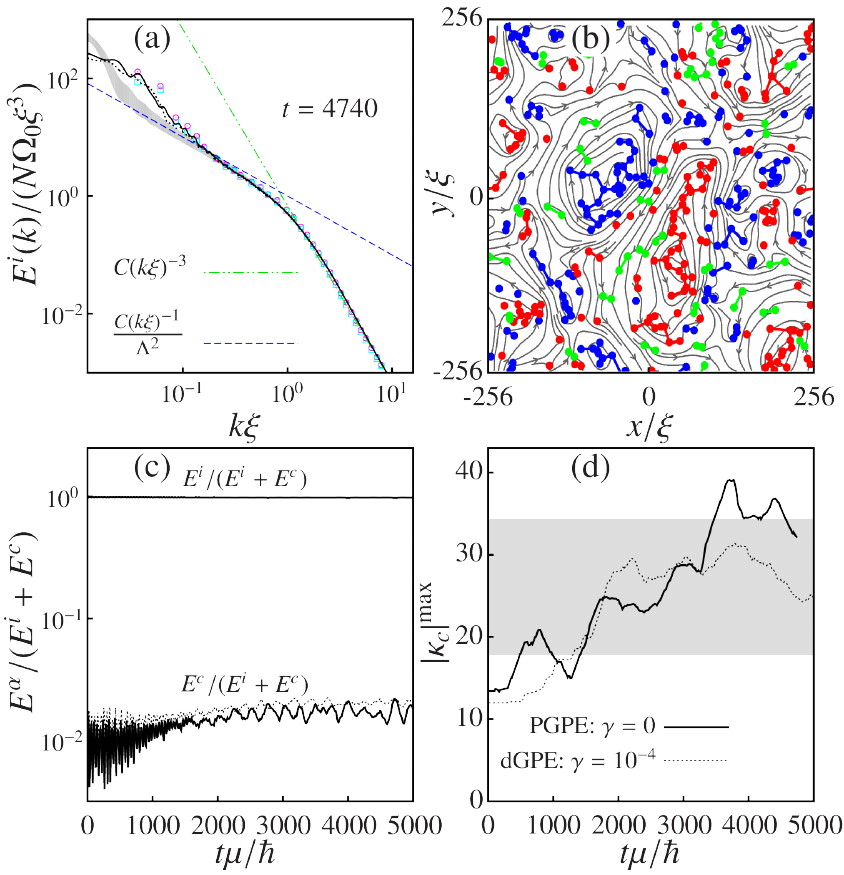}
\caption{Dynamical PGPE evolution of a non-equilibrium neutral 384-vortex
configuration (initial point-vortex energy per vortex $\varepsilon(t=0)=6$)
towards an OKC state (see Fig. 2 of the main text). (a): IKE spectrum, compared
to the statistical equilibrium distribution for $\varepsilon=6$ (grey shaded
line, thickness indicates $\pm1$ standard deviation). The solid black line and
magenta circles show the time-averaged (over 500 natural time units
$\mu/\hbar$) IKE spectrum obtained from the PGPE simulation using the
point-vortex spectrum (Eq.  3 of main text) and the GPE, respectively. The
dashed black line and cyan squares show the equivalent quantities from the dGPE
simulation at the same time. (b): RCA decomposition (see main text) of the
wavefunction in the PGPE simulation at time $t = 4740\mu/\hbar$: the emerging
OKC is clearly visible in the center of the field of view.  (c): Comparison of
the ratios of the total incompressible ($E^i$) and compressible ($E^c$) kinetic
energies to the total kinetic energy $E^i+E^c$ in the PGPE (solid lines) and
dGPE (dashed lines) simulations. In both cases, the initial compressible
kinetic energy is extremely small and fluctuations due to numerical noise at
high-$k$ are visible in the spectrum. At later times the compressible kinetic
energy begins to grow, with faster growth occurring in the PGPE case. However,
in both cases the compressible energy constitutes well below 5\% of the total
kinetic energy throughout the simulation. (d): Time-averaged absolute
charge of the largest vortex cluster, $|\kappa_c|^{\rm max}$, for the PGPE
(solid line) and dGPE (dashed line) simulations. Although the exact dynamics
are different due to the chaotic nature of the vortex trajectories, both
simulations show $|\kappa_c|^{\rm max}$ clearly moving toward the values
expected in statistical equilibrium (shaded area). }
\label{fig:pgpe}
\end{figure}

The dGPE, with $\gamma$ treated as a more-or-less phenomenological parameter,
has been widely-used as a description of finite-temperature Bose-Einstein
condensates. Viewed phenomenologically, the dGPE's key advantage is that it
incorporates some of the dissipative physics lying beyond the zero-temperature
GPE, while remaining computationally tractable even for large, complex systems
\cite{Reeves12a}. Within such a phenomenological treatment one can also attempt
to heuristically include other effects by adjusting $\gamma$; for example in
the case of the 2D dGPE considered here one might expect that the effects of
coupling to compressible dynamics in the tightly-trapped $z$-direction could be
phenomenologically captured by a higher effective damping rate $\gamma$.
However, we emphasize that within the $c$-field approach $\gamma$ can be
calculated \emph{a priori} from experimental parameters, and typically has a
value of order $10^{-4}$ \cite{Bradley2012a}. In this context, the dGPE
simulations presented in our letter go beyond a phenomenological description.
Indeed, recent works have shown that the dGPE can still give a qualitatively
accurate picture of vortex dynamics in oblate-geometry persistent current
formation experiments even at considerably high temperatures compared to $T_c$
\cite{neely_etal_prl_2013, rooney_etal_pra_2013}.

The dynamical effects of the dissipation in the dGPE in the presence of quantum
vortices are twofold: Firstly, the dissipation introduces a direct correction
to the equations of motion for quantum vortices in a completely homogeneous
background \cite{tornkvist_shroder_prl_1997}, introducing a velocity correction
for each vortex proportional to $\gamma$ times the original (Hamiltonian)
velocity.  Secondly, the dissipation suppresses compressible energy at high
wavenumbers $k$. Indeed, this effect is in some ways analogous to the effect of
viscosity in the classical Navier-Stokes equations \cite{Bradley2012a}. The
appearance of dissipative effects predominantly at wavenumbers $k>\xi^{-1}$ was
confirmed numerically in simulations of beyond-mean-field dynamics using the
Hartree-Fock-Bogoliubov description~\cite{kobayashi_tsubota_prl_2006}.
For the small value of $\gamma = 10^{-4}$ used in our simulations the first of
these effects is expected to be negligible over the integration time we
consider ($t< 10^4 \mu / \hbar$). Thus, we expect damping to leave the vortex
dynamics largely unaffected, while supressing sound energy at high $k$, and we
expect our dGPE results for the vortex degrees of freedom to correspond closely to
the Hamiltonian case.

\subsection{B. Relation to Hamiltonian (projected Gross--Pitaevskii) model}
While the dissipative finite-temperature effects captured by the dGPE are
likely to be present to some degree in any experimental realization of
dynamical OKC formation, it is nonetheless interesting to consider what happens
in the Hamiltonian case $\gamma = 0$ numerically. Doing so provides
confirmation that the vortex dynamics remain quantitatively similar in their
statistical properties, and that dynamic OKC formation remains possible, in the
limit of zero-temperature. 

However, simulating the Hamiltonian evolution in a large, highly turbulent
system is a significant numerical challenge.  In the absence of damping,
compressible energy transferred to high $k$ by the turbulent dynamics is not
dissipated and must continue to be spatially and temporally resolved by the
numerical method.  This numerical challenge is similar to that of simulating
condensate formation and inverse energy cascade in two-dimensional classical
fluids \cite{mcwilliams_jfm_1990, Chertkov2007a, xiao_etal_jfm_2009,
chan_etal_pre_2012}. As the central numerical difficulty lies in the aliasing
of high-$k$ modes by the pseudospectral method \cite{boyd}, the projected GPE
(PGPE) \cite{Davis2001b, Davis2006a, Simula2006a} is the appropriate framework
for long-time simulations of Hamiltonian turbulence in a BEC.  Within a
$c$-field framework, the projector in the PGPE formally arises from quantum
mechanical considerations;  however, it is also
directly connected to dealiasing procedures used in turbulent Navier--Stokes
simulations (often in conjunction with a phenomenological hyperviscosity)
\cite{boyd, Blakie08b, shukla_etal_njp_2013}.

Due to the large amount of computational time necessary for Hamiltonian
simulations \footnote{Over $10^4$ CPU-hours on IBM Power 7 architecture} we
have integrated the equations of motion up to the time $t \simeq 5000 \mu /
\hbar$, by which time the onset of OKC is clear in several measures.
\figreftfull{fig:pgpe}{} shows our results for the case $\varepsilon(0)=6$,
obtained using a Fourier pseudospectral method for the undamped PGPE on a
$4096^2$-point spatial grid with adaptive $4$th--$5$th order Runge-Kutta
timestepping (relative error tolerance $10^{-6}$). See also the movies of dGPE
and PGPE evolution accompanying this supplemental material. These results
confirm the predictions of the dGPE [computed on a spatial grid of $2048^2$
points without a projector, with the same timestepping scheme], illustrating
that dynamical OKC formation also occurs in the Hamiltonian case.


\begin{thebibliography}{69}
\bibitem [{sam()}]{sampling}%
  \BibitemOpen
  \href@noop {} {}\bibinfo {note} {We use energy tolerance $\pm{\rm
  max}(0.05\varepsilon, 5\times10^{-3})$. Ensemble sizes in Fig. 1 are $10^2$
  [magenta circles in (c)], $10^4$ [other curves in (c), (d)], $10^7$
  [structure function in (a)], and $10^3$ [other curves in (a),
  (b)].}\BibitemShut {Stop}%
\bibitem [{okc()}]{okcV}%
  \BibitemOpen
  \href@noop {} {}\bibinfo {note} {Note that the OKC is not composed of the
  same vortices at all times; the condensate gradually exchanges vortices with
  the uncorrelated background.}\BibitemShut {Stop}%
\end{thebibliography}
\end{document}